\documentclass[aps,nofootinbib,superscriptaddress, showpacs,preprintnumbers, nofootinbibt,twocolumn]{revtex4-2}
\usepackage{newunicodechar}
\newunicodechar{ℱ}{\mathcal{F}}

\usepackage{epsfig}
\usepackage{amsmath}
\usepackage{multirow}
\usepackage{subcaption}
\usepackage{eurosym}
\usepackage{dcolumn}
\usepackage{bm}
\usepackage{enumerate}
\usepackage{float}
\usepackage{epstopdf}
\usepackage{amsmath}
\usepackage{bm}
\usepackage{amsfonts}
\usepackage{amssymb}
\usepackage{graphicx}
\usepackage{alphalph,mathtools}
\usepackage{etoolbox}
\usepackage{color}
\usepackage{booktabs}
\usepackage{hyperref}
\hypersetup{colorlinks,citecolor=blue}
\usepackage{footnote}
\usepackage{makecell,tabularx}

\usepackage{graphicx} 

\def\be{\begin{equation}}
\def\ee{\end{equation}}
\def\bea{\begin{eqnarray}}
\def\eea{\end{eqnarray}}

\begin{document}

\title{Thermodynamic stability of the new black hole solution in $f(R,G,T)$ gravity} 

\author{Aniruddha Ghosh}
 \email{ruddha.g@gmail.com}
 \affiliation{%
 Department of Mathematics, Indian Institute of Engineering Science and Technology, Shibpur, Howrah-711103, India.\\ 
}%
\author{Ujjal Debnath}%
 \email{ujjaldebnath@gmail.com}
\affiliation{%
 Department of Mathematics, Indian Institute of Engineering Science and Technology, Shibpur, Howrah-711103, India.\\ 
}%

\begin{abstract}
In this literature, we derive an approximate black hole metric within the \( f(R, G, T) \) gravity framework, where the black hole is surrounded by an anisotropic fluid as the matter source. This approach allows us to study the influence of anisotropic pressures on the black hole structure. We then investigate the thermodynamic properties of the approximate solution, considering various types of matter fields. We analyze the thermodynamic quantities such as Wald's entropy, Hawking temperature, and specific heat for each case. After examining the specific heat, we find that the black holes are thermodynamically stable, with positive specific heat values indicating stability. This result suggests that the black holes under these conditions do not undergo thermal instability, ensuring their long-term viability.\\

\end{abstract}

\maketitle
\section{Introduction}\label{sec1}
Observational data strongly suggest that our universe is expanding \cite{I1}. This conclusion is primarily based on the redshift of light from distant galaxies, which indicates that they are moving away from us. More precise measurements, such as those involving Type Ia supernovae\cite{I1}, the Cosmic Microwave Background (CMB), and large-scale structure, have revealed that not only is the universe expanding, but this expansion is accelerating over time. To explain this accelerated expansion, several theoretical approaches have been proposed. The most widely accepted explanation within the framework of General Relativity is the presence of dark energy \cite{I2,I3,I4,I5} — an unknown form of energy that permeates space and exerts negative pressure. However, the nature of dark energy remains one of the biggest mysteries in modern cosmology. As an alternative to invoking dark energy, some physicists have proposed modifications to Einstein's theory of General Relativity, an approach broadly known as modified gravity \cite{basak2025accretion,mukherjee2024accretion}. These theories attempt to explain cosmic acceleration by changing the laws of gravity on large (cosmological) scales.
\par
\vspace{0.25cm}
To go beyond General Relativity, many alternative theories of gravity are constructed by extending the Einstein-Hilbert action with additional terms. These modifications introduce new dynamics that can address unresolved issues in cosmology, such as the accelerated expansion of the universe. Various approaches have been explored in this context, including theories like \( f(R) \) gravity~\cite{I6, I7, I8}, which generalizes the Ricci scalar; \( f(G) \) gravity~\cite{I9, I10}, which involves the Gauss--Bonnet invariant; $f(R, G)$ \cite{I22, I23}gravity and \( f(\mathcal{T}) \) gravity~\cite{I11, I12, I13, I14}, based on the torsion scalar in teleparallelism. More complex formulations, such as \( f(\mathcal{T}, T_G) \) gravity~\cite{I15,I16}, combine torsional and Gauss--Bonnet contributions and $f(P)$ gravity \cite{I17} . Other notable examples include Weyl gravity~\cite{I18, I19}, which incorporates conformal symmetry, and Lovelock gravity~\cite{I20,I21}.However, studying these modified gravity theories poses significant difficulties, primarily due to the higher-order derivatives involved in their equations of motion. Unlike General Relativity, which leads to second-order differential equations, many of these models yield field equations of fourth order or even higher. In our specific case, the resulting equation of motion is a non-linear differential equation of fourth order, making exact solutions hard to obtain.
\par
\vspace{0.25cm}
Another novel class of modified gravity theories involves higher-order curvature invariants and is represented by the \( f(R, G, T) \) framework \cite{I24}. In this formulation, \( R \) denotes the Ricci scalar, \( G \) is the Gauss-Bonnet invariant, and \( T \) represents the trace of the energy-momentum tensor. A distinctive feature of this theory is that the energy-momentum tensor is generally not conserved. This non-conservation arises due to the explicit coupling between matter and geometry, which leads to a non-zero covariant divergence of the energy-momentum tensor. 
\par
\vspace{0.25cm}
In this paper, we construct black hole solutions within the framework of \( f(R, G, T) \) gravity, considering a surrounding anisotropic fluid. We explore the thermodynamic properties of these black holes in the presence of various matter fields. The field equations in this modified theory are highly complex, consisting of fourth-order nonlinear differential equations. To address this challenge, we propose a general method for obtaining approximate solutions under such conditions. After deriving the approximated solutions, we analyze key thermodynamic quantities such as the Hawking temperature \textbf{T}, Wald entropy \( S \), and specific heat \( C \) as functions of the event horizon radius \( r_h \). Our approach provides insights into the thermodynamic behavior of black holes in higher-order modified gravity theories.
Similar studies have been conducted in other modified gravity  models, including \( f(R) \) gravity\cite{I25,I26,I27}, \( f(G) \) gravity\cite{I28,I29}, \(f(\mathcal{T}) \) gravity\cite{I30,I31,I32}, and \( f(P) \) gravity\cite{I17}, among others.
\par
\vspace{0.20cm}
The structure of this paper is as follows: In Section \ref{sec2}, we formulate the field equations for \( f(R, G, T) \) gravity, considering an anisotropic fluid as a matter source. We then derive the corresponding differential equations governing the black hole metric. Furthermore, we present a method to obtain approximate solutions to these complex equations and explicitly construct the black hole metric.
In Section \ref{sec3}, we analyze the thermodynamic properties of the obtained solutions. Specifically, we compute the Wald entropy, Hawking temperature, and specific heat for black holes . Additionally, we identify solutions that are thermodynamically stable—i.e., those with positive specific heat.

\section{FIELD EQUATION OF $f(R,G,T)$ GRAVITY}\label{sec2}
In this section, we formulate the Einstein field equations within the \( f(R, G, T) \) modified gravity framework, starting with the corresponding action for this theory.
\begin{equation} \label{1}
      I=\int \sqrt{-g}\left(\frac{1}{2k^2}f(R,G,T)+L_{m}\right)\ d^4x  
    \end{equation}
    where \( f(R, G, T) \) is an arbitrary function of the Ricci scalar \( R \), the Gauss-Bonnet invariant \( G \), and the trace \( T \) of the stress-energy tensor of the matter. Also, \( L_m \) is the matter Lagrangian, \( g = |g_{\mu\nu}| \), and \( \kappa^2 = 8\pi G_N \) (choosing \( c = 1 \)). The stress-energy tensor of the matter is defined as \cite{F1}
\begin{equation} \label{2}
T_{\mu\nu} = -\frac{2}{\sqrt{-g}} \frac{\delta \left( \sqrt{-g} L_m \right)}{\delta g^{\mu\nu}}
\end{equation}
The trace of the stress-energy tensor is \( T = g^{\mu\nu} T_{\mu\nu} \), and assuming \( L_m \) depends only on \( g_{\mu\nu} \), we get \( T_{\mu\nu} = g_{\mu\nu} L_m - 2 \frac{\partial L_m}{\partial g^{\mu\nu}} \).
Varying the action (\ref{1}) leads to the following integral.
    \begin{equation} \label{3}
\begin{split}
      \delta I =& \frac{1}{2\kappa^2} \int \Big[ f_R \delta R + f_G \delta G + f_T \delta T + f \delta \left( \sqrt{-g} \right) \\
      &+ \frac{2\kappa^2}{\sqrt{-g}} \delta \left( \mathcal{L}_m \sqrt{-g} \right) \Big]\sqrt{-g}d^4x
\end{split}
\end{equation}
The Ricci scalar \( R \) and Gauss–Bonnet invariant \( G \) are as follows:

\[
R = g^{\mu\nu} R_{\mu\nu}
\]

The Gauss–Bonnet invariant \( G \) is given by:

\[
G = R^2 - 4 R_{\mu\nu} R^{\mu\nu} + R_{\mu\nu\xi\eta} R^{\mu\nu\xi\eta}
\]

The partial derivatives of the function \( f \) with respect to the Ricci scalar \( R \), the Gauss–Bonnet invariant \( G \), and some other quantity \( T \) are given by:

\[
f_R = \frac{\partial f}{\partial R}, \quad f_G = \frac{\partial f}{\partial G}, \quad f_T = \frac{\partial f}{\partial T}
\]
The variations of \( \sqrt{-g} \), \( R \), \( R_{\mu\nu} \), \( G \), and \( T \) are as follows:
\begin{equation} \label{4}
 \delta \sqrt{-g} = -\frac{1}{2} \sqrt{-g} \, g^{\mu\nu} \delta g_{\mu\nu}
   \end{equation}
\begin{equation}\label{5}
\begin{split}
  f_R \delta R =& (f_R \frac{\partial R}{\partial R^{\mu \sigma\rho\lambda}} R_{\nu}^{\sigma\rho\lambda}-2\nabla^\alpha \nabla^\beta \frac{\partial R}{\partial R^{\mu \alpha\beta\nu}}f_{R})\delta g^{\mu\nu}
   \\
  =&( R_{\mu\nu}  + g_{\mu\nu} \nabla^2  - \nabla_\mu \nabla_\nu )f_{R}\delta g^{\mu\nu}
  \end{split}
\end{equation}

\begin{equation}\label{6}
  f_{G} \delta G = (f_G \frac{\partial G}{\partial R^{\mu \sigma\rho\lambda}} R_{\nu}^{\sigma\rho\lambda}-2\nabla^\alpha \nabla^\beta \frac{\partial G}{\partial R^{\mu \alpha\beta\nu}}f_{G})\delta g^{\mu\nu}\\
\end{equation}
\begin{equation}\label{7}
   \delta T = T_{\mu\nu} \delta g^{\mu\nu} + \Theta_{\mu\nu} \delta g^{\mu\nu}
 \end{equation}
\begin{equation}\label{8}
\Theta_{\mu\nu} = g^{\alpha\beta} \frac{\partial T_{\alpha\beta}}{\partial g_{\mu\nu}}
\end{equation}
where 
\begin{equation}\label{9}
\begin{split}
    \frac{\partial G}{\partial R^{\mu \sigma\rho\nu}}=&R(g_{\mu \rho}g_{\mu \sigma}-g_{\mu \lambda}g_{\rho \sigma})-\\
    &2(-g_{\rho \sigma}R_{\lambda \mu}+g_{\mu \rho}R_{\lambda \sigma}+g_{\lambda \sigma}R_{\mu \rho}-g_{\lambda \mu}R_{\rho \sigma})+2R_{\mu \sigma\rho\nu}
\end{split}
\end{equation}
Now applying  $\delta I=0$,we obtain the field equation
\begin{equation}\label{10}
\begin{split}
&(R_{\mu\nu} + g_{\mu\nu} \nabla^2 - \nabla_\mu \nabla_\nu )f_R - \frac{1}{2} f g_{\mu\nu}+\\ 
& (f_G \frac{\partial G}{\partial R^{\mu \sigma\rho\lambda}} R_{\nu}^{\sigma\rho\lambda}-2\nabla^\alpha \nabla^\beta \frac{\partial G}{\partial R^{\mu \alpha\beta\nu}}f_{G})= \kappa^2 T_{\mu\nu} - (T_{\mu\nu} + \Theta_{\mu\nu}) f_T
\end{split}
\end{equation}
Taking the covariant divergence of Eq. (\ref{10}), we obtain:\cite{I23,F2,F3,F4,F5,F6,F7,F8,F9,F10}

\begin{equation}\label{11}
\nabla^\mu T_{\mu\nu} = \frac{f_T}{\kappa^2-f_T}\Big[ \left( T_{\mu\nu} + \Theta_{\mu\nu} \right) \nabla^\mu \ln f_T + \nabla^\mu \Theta_{\mu\nu} - \frac{1}{2} g_{\mu\nu} \nabla^\mu T\Big]
\end{equation}
The expression is independent of \( f_R \) and \( f_G \), but since \( f \) depends on \( T \), \( f_T \neq 0 \), leading to a violation of energy conservation. Also, Eq. (\ref{8}) is written as follows:

\begin{equation}\label{12}
\Theta_{\mu\nu} = -2 T_{\mu\nu} + g_{\mu\nu} L_m - 2 g^{\alpha\beta} \frac{\partial^2 L_m}{\partial g_{\mu\nu} \partial g_{\alpha\beta}}
\end{equation}
The energy-momentum tensor of the surrounding field 
 defined as \cite{F11}

\begin{equation}\label{13}
T^{\mu\nu} = (\rho + p_t) u^\mu u^\nu + p_t g_{\mu\nu}+(p_r-p_t)v^\mu v^\nu
\end{equation}
Here, \( p_r \) and \( p_t \) denote the radial and tangential pressures of the anisotropic fluid, respectively.
From Eq.(\ref{11}), we obtain the nonconservation equation
\begin{equation}\label{14}
   - 2(p_r+2p_t-3\rho)\dot{f_T} -f(\dot{p_r}+\dot{2p_t})+3(3f+2k^2)\dot{\rho}=0
\end{equation}
From equation (\ref{14}), we obtain the form of the function $f(R, G, T)$
\begin{equation}\label{15}
    f(R,G,T)=c_1 T+\phi(R,G)
\end{equation}
where $c_1$ and $\phi(R,G)$ are arbitrary constant and function respectively.
Here we choose the function $f(R,G,T)$ of the form
\begin{equation}\label{16}
   f(R,G,T)=\beta_1 R+\beta_2G^2+\beta_3 T 
\end{equation}
Therefore, the field equation (\ref{9}) became
\begin{equation}\label{17}
  \begin{split}
&\tilde{G_{\mu\nu}}=(R_{\mu\nu} + g_{\mu\nu} \nabla^2 - \nabla_\mu \nabla_\nu )\beta_1  - \frac{1}{2} (\beta_1 R+\beta_2G^2)g_{\mu\nu}+\\ 
& (2G\beta_2 \frac{\partial G}{\partial R^{\mu \sigma\rho\lambda}} R_{\nu}^{\sigma\rho\lambda}-2\nabla^\alpha \nabla^\beta \frac{\partial G}{\partial R^{\mu \alpha\beta\nu}}2G\beta_2)\\
&= \kappa^2 T_{\mu\nu} - (T_{\mu\nu} + \Theta_{\mu\nu}) f_T+\frac{1}{2} \beta_3 Tg_{\mu\nu}=\tilde{T_{\mu\nu}}
\end{split}
\end{equation}  
Here, we assume the static, spherically symmetric black hole solutions described by a single function \( \psi(r) \) correspond to the metric:
\begin{equation}\label{18}
ds^2 = -\psi(r) \, dt^2 + \frac{dr^2}{\psi(r)} + r^2 \left( d\theta^2 + \sin^2\theta \, d\phi^2 \right)
\end{equation}
And also we choose $p_r$ and $p_t$ as follows
\begin{equation}\label{19}
\begin{split}
  &  p_r=-\rho\\
& p_t=\frac{1}{2}\rho(3\omega+1)
\end{split}
\end{equation}
 where \( \rho \) is the energy density and \( \omega \) is a state parameter.For an anisotropic fluid, $w \neq -1$.
 The four-velocity \( u^\mu \) and radial vector $v^{\mu}$ satisfies the conditions \( u^\mu u_\mu = -1 \), $ v^{\mu}v_{\nu}=1$ ,$v^{\mu}u_{\mu}=0$ and \( u^\mu \nabla _\nu u_\mu = 0 \).
In the context of static and spherically symmetric spacetimes, the line element is described by a single metric function \( \psi(r) \). However, when matter is present, the gravitational field equations generally involve not only the metric components but also the matter content, typically represented by the energy density \( \rho(r) \). 

Consequently, the problem reduces to determining two unknown functions: the metric function \( \psi(r) \), which characterizes the geometry of spacetime, and the energy density \( \rho(r) \), which encodes the matter distribution. Solving for both simultaneously requires a system of two independent differential equations, typically obtained from the modified field equations of the theory under consideration.

However, if one of the functions is specified—such as assuming a known form for \( \rho(r) \) based on a physical model, equation of state, or observational input—then only a single independent differential equation is needed to solve for the remaining unknown, namely \( \psi(r) \). 
 Taking the \( r \)-component of Eq. \ref{11}, we obtain the following equation
\begin{equation}\label{20}
-6 (1 + w) (k^2 + \beta_3) \rho(r) + \left(-2 k^2 + (-3 + 5w)\beta_3 \right) r \, \rho'(r) = 0
\end{equation}
Solving the equation (\ref{20}), we obtain $\rho(r)$
\begin{equation}\label{21}
    \rho(r) = C_1 \, r^{-\frac{6 (1 + w) (k^2 + \beta_3)}{2 k^2 + (3 - 5w) \beta_3}}
\end{equation}
Where $C_1$ is an arbitrary constant. Now, we need to find a single unknown, the metric function \( \psi(r) \). We consider the \( t \)-component of the modified field equations (\ref{17}). Then, we obtain the required equation as follows:
\begin{widetext}
\begin{equation}\label{22}
\begin{split}
& r^{5 + \frac{6 (1 + w) (k^2 + \beta_3)}{-2 k^2 + (-3 + 5 w) \beta_3}} 
(2 k^2 + (3 - 5 w) \beta_3) C_1 + \frac{1}{r} \cdot 2 \Bigg[ \\
& \quad 64 \beta_2 \psi(r)^3 \left(3 \psi''(r) - r \psi^{(3)}(r)\right) + r \Big( 
-200 r \beta_2 (\psi'(r))^4 
+ 8 r \beta_2 (\psi''(r))^2 (-1 + r^2 + 3 r^2 \psi''(r)) \\
& \quad + 16 r \beta_2 (\psi'(r))^2 \left(\psi''(r) (13 + r^2 + 3 r^2 \psi''(r)) - 2 r \psi^{(3)}(r)\right) \\
& \quad + 8 \beta_2 (\psi'(r))^3 \left(8 - 4 r^2 + 8 r^2 \psi''(r) + 3 r^3 \psi^{(3)}(r)\right) \\
& \quad + \psi'(r) \left(r^4 \beta_1 + 32 \beta_2 \psi''(r) (-2 + r^2 + r^2 \psi''(r)) - 8 r^3 \beta_2 (1 + 6 \psi''(r)) \psi^{(3)}(r)\right) \Big) \\
& \quad + 8 \beta_2 \psi(r)^2 \Big( 
24 (\psi'(r))^2 - 33 r^2 (\psi''(r))^2 + 4 r \psi^{(3)}(r) (4 - r^2 + 3 r^3 \psi^{(3)}(r)) + 2 r^4 \psi^{(4)}(r) \\
& \quad + 4 \psi''(r) (-12 + r^2 + 4 r^3 \psi^{(3)}(r) + 3 r^4 \psi^{(4)}(r)) \\
& \quad + 4 r \psi'(r) \left(4 \psi''(r) + r (-9 \psi^{(3)}(r) + 2 r \psi^{(4)}(r)) \right) \Big) \\
& \quad + \psi(r) \Big( 
r^4 \beta_1 + 8 \beta_2 \Big(
40 r (\psi'(r))^3 + r^2 (34 + 5 r^2) (\psi''(r))^2 + 15 r^4 (\psi''(r))^3 \\
& \quad + 4 r \psi^{(3)}(r)(-2 + r^2 - 3 r^3 \psi^{(3)}(r)) - 2 r^4 \psi^{(4)}(r) \\
& \quad - 4 \psi''(r)(-6 + r^2 + 4 r^3 \psi^{(3)}(r) + 3 r^4 \psi^{(4)}(r)) \\
& \quad + 2 (\psi'(r))^2 (-12 + 2 r^2 - 83 r^2 \psi''(r) + 20 r^3 \psi^{(3)}(r) + 3 r^4 \psi^{(4)}(r)) \\
& \quad + r \psi'(r) \left(
56 r^2 (\psi''(r))^2 + 2 \psi''(r)(-4 - 8 r^2 + 33 r^3 \psi^{(3)}(r)) + r (9 (4 + r^2) \psi^{(3)}(r) - 8 r \psi^{(4)}(r)) 
\right) \Big) \Big) 
\Bigg] \\
& - 2 r^3 \beta_1=0
\end{split}
\end{equation}
\end{widetext}
Even now, solving equation (\ref{22}) remains a complex task. Therefore, we propose an approximate method to solve equation (\ref{22}). First, we set \( \beta_2 = 0 \) in equation (\ref{22}) to reduce the equation to one that only involves \( R \) and \( T \) (\ref{23}).
\begin{equation}\label{23}
\begin{split}
&r^{5 + \frac{6 (1 + w)(k^2 + \beta_3)}{-2 k^2 + (-3 + 5w)\beta_3}} 
(2k^2 + (3 - 5w)\beta_3) C_1 
\\
&+ \frac{2 \left( r^4 \beta_1 \psi(r) + r^5 \beta_1 \psi'(r) \right)}{r} 
-2 r^3 \beta_1=0
\end{split}
\end{equation}   
 By solving the reduced differential equation (\ref{23}), we obtain the metric function for \( \beta_2 = 0 \).
\begin{equation}\label{24}
    \psi(r) = 
1 + 
\frac{
r^{2 + \frac{6 (1 + w)(k^2 + \beta_3)}{-2 k^2 + (-3 + 5w)\beta_3}} 
\left(2k^2 + (3 - 5w)\beta_3 \right)^2 C_1
}{
6 \beta_1 \left(2 k^2 w + (-1 + 7w)\beta_3 \right)
}
+ \frac{C_2}{r}
\end{equation}
Where $C_1, C_2$ are arbitrary constants.We now rewrite equation (\ref{24}) in the following form:
\begin{equation}\label{25}
\psi(r) = 1 + j \, r^{\alpha} + \frac{C_2}{r}
\end{equation}
Where $$j=\frac{\left(2k^2 + (3 - 5w)\beta_3 \right)^2 C_1}{6 \beta_1 \left(2 k^2 w + (-1 + 7w)\beta_3 \right)} 
$$and $$\alpha={2 + \frac{6 (1 + w)(k^2 + \beta_3)}{-2 k^2 + (-3 + 5w)\beta_3}}$$
\\

We aim to solve equation (\ref{22}) using a first-order approximation in \( \beta_{2} \). For this purpose, we consider the solution of the differential equation (\ref{22}) in the following form:
\begin{equation}\label{26}
\psi(r) = 1 + j \, r^{\alpha} - \frac{1}{r} + \beta_2  t(r)
\end{equation}

\begin{figure}[H]
    \centering
    \includegraphics[width=1\linewidth]{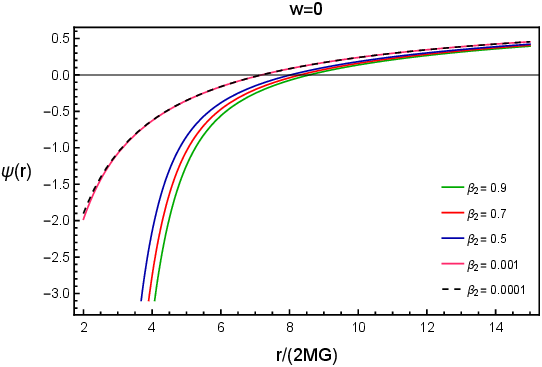}
    \caption{We plot $\psi(r)$ for various $\beta_2$ values for $w=0$}
    \label{f1}
    \end{figure}

\begin{figure}[H]
    \centering
    \includegraphics[width=1\linewidth]{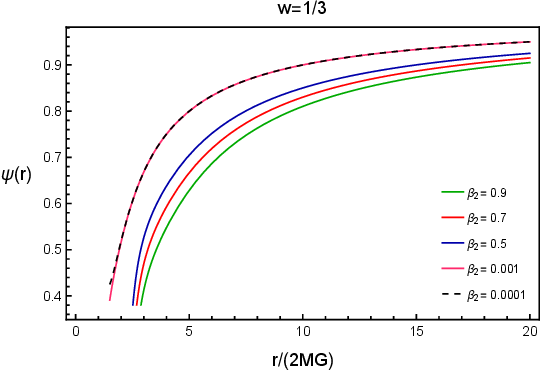}
    \caption{We plot $\psi(r)$ for various $\beta_2$ values for $w=1/3$}
    \label{f2}
    \end{figure}

Now, substituting equation (\ref{26}) into equation (\ref{22}) and collecting the terms linear in \( \beta_2 \), we obtain the differential equation in terms of \( t(r) \). By solving the differential equation, we obtain the function \( t(r) \) (\ref{27}). Also we consider $C_1=1, C_2=-1$ and Einstein gravitational constant \( k = 1 \).
\begin{widetext}
\begin{equation}\label{27}
\begin{split}
t(r)=&\frac{520}{r^{10} \beta_1} - \frac{576}{r^9 \beta_1} - \frac{264}{r^7 \beta_1} + \frac{288}{r^6 \beta_1} - \frac{128 j^3 r^{-7 + 3 \alpha}}{\beta_1} + \frac{8352 j r^{-9 + \alpha}}{(-8 + \alpha) \beta_1} - \frac{4096 j r^{-8 + \alpha}}{(-8 + \alpha) \beta_1} - \frac{3456 j r^{-6 + \alpha}}{(-8 + \alpha) \beta_1} \\
&+ \frac{1280 j r^{-5 + \alpha}}{(-8 + \alpha) \beta_1} - \frac{112 j^3 r^{-7 + 3 \alpha} \alpha}{3 \beta_1} - \frac{64 j^3 r^{-6 + 3 \alpha} \alpha}{\beta_1} - \frac{64 j^3 r^{-4 + 3 \alpha} \alpha}{\beta_1} + \frac{688 j r^{-9 + \alpha} \alpha}{(-8 + \alpha) \beta_1} \\
&- \frac{512 j r^{-8 + \alpha} \alpha}{(-8 + \alpha) \beta_1} + \frac{2224 j r^{-6 + \alpha} \alpha}{(-8 + \alpha) \beta_1} - \frac{2336 j r^{-5 + \alpha} \alpha}{(-8 + \alpha) \beta_1} - \frac{168 j^3 r^{-7 + 3 \alpha} \alpha^2}{\beta_1} - \frac{64 j^3 r^{-6 + 3 \alpha} \alpha^2}{\beta_1} \\
&+ \frac{152 j^3 r^{-4 + 3 \alpha} \alpha^2}{\beta_1} + \frac{8392 j r^{-9 + \alpha} \alpha^2}{(-8 + \alpha) \beta_1} - \frac{9216 j r^{-8 + \alpha} \alpha^2}{(-8 + \alpha) \beta_1} - \frac{1184 j r^{-6 + \alpha} \alpha^2}{(-8 + \alpha) \beta_1} + \frac{1296 j r^{-5 + \alpha} \alpha^2}{(-8 + \alpha) \beta_1} \\
&+ \frac{1648 j^3 r^{-7 + 3 \alpha} \alpha^3}{3 \beta_1} + \frac{208 j^3 r^{-6 + 3 \alpha} \alpha^3}{\beta_1} - \frac{48 j^3 r^{-4 + 3 \alpha} \alpha^3}{\beta_1} - \frac{2488 j r^{-9 + \alpha} \alpha^3}{(-8 + \alpha) \beta_1} + \frac{2576 j r^{-8 + \alpha} \alpha^3}{(-8 + \alpha) \beta_1} \\
&+ \frac{248 j r^{-6 + \alpha} \alpha^3}{(-8 + \alpha) \beta_1} - \frac{256 j r^{-5 + \alpha} \alpha^3}{(-8 + \alpha) \beta_1} - \frac{912 j^3 r^{-7 + 3 \alpha} \alpha^4}{\beta_1} + \frac{496 j^3 r^{-6 + 3 \alpha} \alpha^4}{\beta_1} + \frac{176 j r^{-9 + \alpha} \alpha^4}{(-8 + \alpha) \beta_1} \\
&- \frac{176 j r^{-8 + \alpha} \alpha^4}{(-8 + \alpha) \beta_1} - \frac{16 j r^{-6 + \alpha} \alpha^4}{(-8 + \alpha) \beta_1} + \frac{16 j r^{-5 + \alpha} \alpha^4}{(-8 + \alpha) \beta_1} + \frac{360 j^3 r^{-7 + 3 \alpha} \alpha^5}{\beta_1} - \frac{288 j^3 r^{-6 + 3 \alpha} \alpha^5}{\beta_1} \\
&- \frac{4448 j^2 r^{-8 + 2 \alpha}}{(-7 + 2 \alpha) \beta_1} + \frac{896 j^2 r^{-7 + 2 \alpha}}{(-7 + 2 \alpha) \beta_1} + \frac{1120 j^2 r^{-5 + 2 \alpha}}{(-7 + 2 \alpha) \beta_1} - \frac{656 j^2 r^{-8 + 2 \alpha} \alpha}{(-7 + 2 \alpha) \beta_1} \\
&+ \frac{416 j^2 r^{-7 + 2 \alpha} \alpha}{(-7 + 2 \alpha) \beta_1} - \frac{2504 j^2 r^{-5 + 2 \alpha} \alpha}{(-7 + 2 \alpha) \beta_1} + \frac{448 j^2 r^{-4 + 2 \alpha} \alpha}{(-7 + 2 \alpha) \beta_1} - \frac{8584 j^2 r^{-8 + 2 \alpha} \alpha^2}{(-7 + 2 \alpha) \beta_1} \\
&+ \frac{1488 j^2 r^{-7 + 2 \alpha} \alpha^2}{(-7 + 2 \alpha) \beta_1} + \frac{2360 j^2 r^{-5 + 2 \alpha} \alpha^2}{(-7 + 2 \alpha) \beta_1} - \frac{1248 j^2 r^{-4 + 2 \alpha} \alpha^2}{(-7 + 2 \alpha) \beta_1} + \frac{8600 j^2 r^{-8 + 2 \alpha} \alpha^3}{(-7 + 2 \alpha) \beta_1} \\
&- \frac{6304 j^2 r^{-7 + 2 \alpha} \alpha^3}{(-7 + 2 \alpha) \beta_1} - \frac{1000 j^2 r^{-5 + 2 \alpha} \alpha^3}{(-7 + 2 \alpha) \beta_1} + \frac{768 j^2 r^{-4 + 2 \alpha} \alpha^3}{(-7 + 2 \alpha) \beta_1} - \frac{5232 j^2 r^{-8 + 2 \alpha} \alpha^4}{(-7 + 2 \alpha) \beta_1} \\
&+ \frac{5360 j^2 r^{-7 + 2 \alpha} \alpha^4}{(-7 + 2 \alpha) \beta_1} + \frac{144 j^2 r^{-5 + 2 \alpha} \alpha^4}{(-7 + 2 \alpha) \beta_1} - \frac{128 j^2 r^{-4 + 2 \alpha} \alpha^4}{(-7 + 2 \alpha) \beta_1} + \frac{1680 j^2 r^{-8 + 2 \alpha} \alpha^5}{(-7 + 2 \alpha) \beta_1} \\
&- \frac{1728 j^2 r^{-7 + 2 \alpha} \alpha^5}{(-7 + 2 \alpha) \beta_1} - \frac{192 j^2 r^{-8 + 2 \alpha} \alpha^6}{(-7 + 2 \alpha) \beta_1} + \frac{192 j^2 r^{-7 + 2 \alpha} \alpha^6}{(-7 + 2 \alpha) \beta_1} + \frac{320 j^4 r^{-6 + 4 \alpha} \alpha}{(-5 + 4 \alpha) \beta_1} \\
&+ \frac{72 j^4 r^{-6 + 4 \alpha} \alpha^2}{(-5 + 4 \alpha) \beta_1} - \frac{1288 j^4 r^{-6 + 4 \alpha} \alpha^3}{(-5 + 4 \alpha) \beta_1} - \frac{1608 j^4 r^{-6 + 4 \alpha} \alpha^4}{(-5 + 4 \alpha) \beta_1} + \frac{3152 j^4 r^{-6 + 4 \alpha} \alpha^5}{(-5 + 4 \alpha) \beta_1} \\
&- \frac{960 j^4 r^{-6 + 4 \alpha} \alpha^6}{(-5 + 4 \alpha) \beta_1} + \frac{C_3}{r}
\end{split}
\end{equation}
\end{widetext}
Where $$j=\frac{\left(2 + (3 - 5w)\beta_3 \right)^2 }{6 \beta_1 \left(2  w + (-1 + 7w)\beta_3 \right)} 
$$and $$\alpha={2 + \frac{6 (1 + w)(1 + \beta_3)}{-2  + (-3 + 5w)\beta_3}}$$
Therefore, the approximated metric function  is :
\begin{widetext}
\begin{equation}\label{28}
\Large{\psi(r) = 
1 + 
\frac{
r^{2 + \frac{6 (1 + w)(k^2 + \beta_3)}{-2 k^2 + (-3 + 5w)\beta_3}} 
\left(2k^2 + (3 - 5w)\beta_3 \right)^2 
}{
6 \beta_1 \left(2 k^2 w + (-1 + 7w)\beta_3 \right)
}
- \frac{1}{r}+\beta_{2}t(r)} +\mathcal{O}(\beta_{2}^n)
\end{equation}
where , $t(r)$ given by the equation (\ref{27}).
\end{widetext}
From this point onward, we consider equation (\ref{28}) to be the approximate metric function and analyze the thermodynamic properties of the black hole in the regime of small \( \beta_2 \) values.
Figures \ref{f1} and \ref{f2} illustrate the black hole metric in the presence of surrounding matter fields---\( w = 0 \) and  \( w = \tfrac{1}{3} \).

    \section{Black hole thermodynamics}\label{sec3}
In this section, we conduct a detailed analysis of various thermodynamic properties \cite{BHT1,BHT2,BHT3,BHT4} associated with the approximate solutions (\ref{28}). Specifically, we examine key quantities such as the Hawking temperature \cite{BHT5}, Wald entropy \cite{BHT6,BHT7}, and specific heat capacity, which provide important insights into the physical behavior and stability of the Black hole.
  The entropy can be computed using Wald’s prescription. This method provides a general framework for deriving black hole entropy as follows:
  \begin{equation}\label{29}
 S = -2\pi \oint \sqrt{\sigma} \, P^{abcd} \, \epsilon_{ab} \, \epsilon_{cd} \, d^2x
  \end{equation}
    where $ P^{abcd}=\frac{\partial \mathcal{L}}{R^{abcd}}$ and $\mathcal{L}$ in Lagrangian.
  Here, \( \sigma \) denotes the determinant of the induced metric on the horizon, and \( \epsilon_{ab} \) represents the binormal vector to the horizon, normalized such that \( \epsilon_{ab} \epsilon^{ab} = -2 \). We consider a spacetime metric of the form given in equation(\ref{18}), with a spherical horizon located at \( r = r_h \).
   \begin{figure}[H]
    \centering
    \includegraphics[width=1\linewidth]{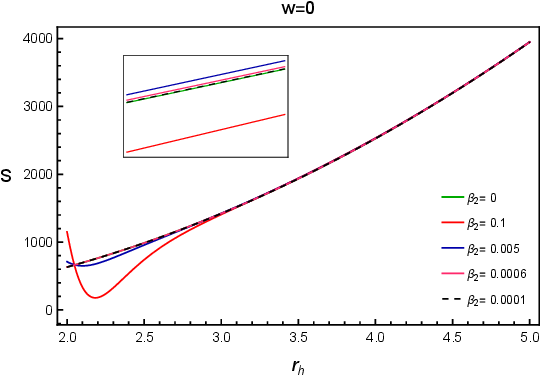}
    \caption{We plot $S$ for various $\beta_2$ values for dust field.}
        \label{f4}
    \end{figure}
   
   \begin{figure}[H]
    \centering
    \includegraphics[width=1\linewidth]{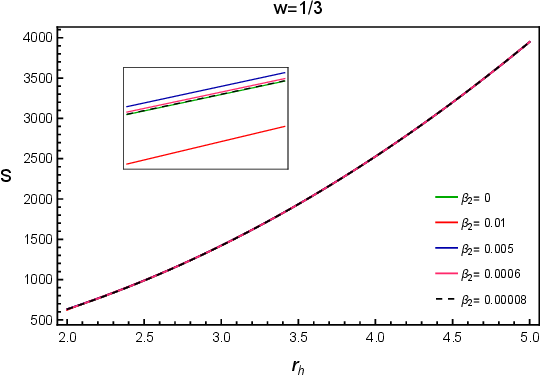}
    \caption{We plot $S$ for various $\beta_2$ values for radiation  field}
        \label{f5}
    \end{figure}
As illustrated in Fig.\ref{f4} and Fig.\ref{f5}, the entropy increases monotonically with the horizon radius. Furthermore, it can be observed that for sufficiently large horizon radius, all curves converge, indicating stable thermodynamic behavior for the black hole.\\

On the other hand, the Hawking temperature of our solutions, as derived in \cite{BHT5}, can be expressed as a function of the horizon radius as follows:
\begin{equation}\label{30}
    \textbf{T} = \frac{\psi'(r_h)}{4\pi}
\end{equation}

       \begin{figure}[H]
    \centering
    \includegraphics[width=1\linewidth]{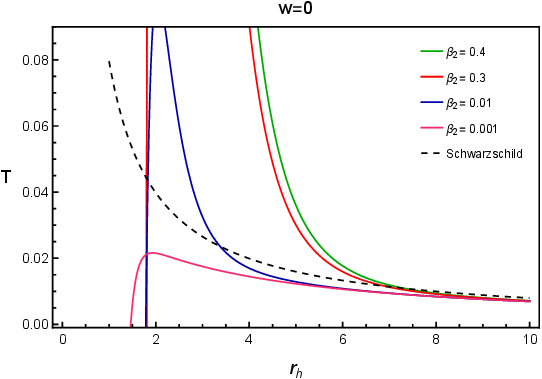}
    \caption{We plot $\textbf{T}$ for various $\beta_2$ values for dust field.}
    \label{f7}
    \end{figure}
   
   \begin{figure}[H]
    \centering
    \includegraphics[width=1\linewidth]{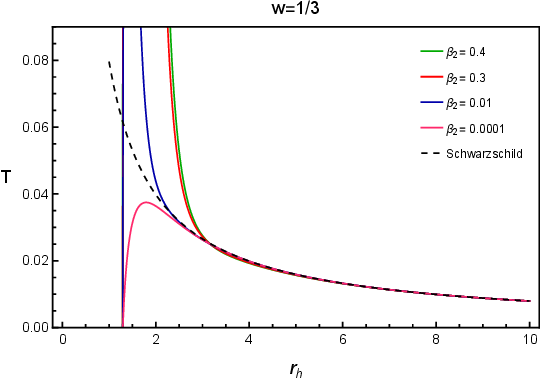}
    \caption{We plot $\textbf{T}$ for various $\beta_2$ values for radiation  field}
    \label{f8}
    \end{figure}
   
    Figures \ref{f7} and \ref{f8} illustrate the behavior of the Hawking temperature as a function of the horizon radius. Initially, the temperature increases, reaching a maximum at a specific horizon size, after which it begins to decrease. Notably, for a large horizon radius, the curves corresponding to different parameter values tend to converge, indicating a universal thermal behavior in this regime.
    Now, we proceed to calculate the specific heat, which we define as follows:
\begin{equation}\label{31}
        C = T \left( \frac{\partial S}{\partial T} \right)
 \end{equation}
 \begin{figure}[H]
    \centering
    \includegraphics[width=1\linewidth]{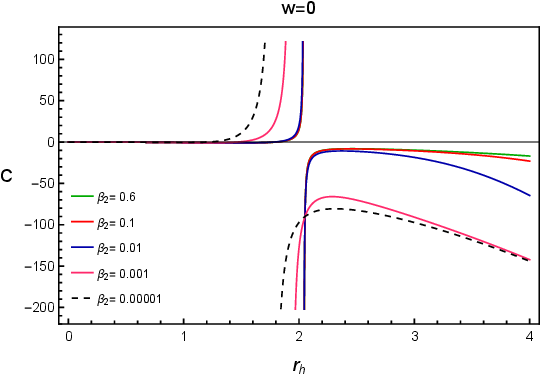}
    \caption{We plot $C$ for various $\beta_2$ values for dust field.}
    \label{f10}
    \end{figure}
   
   \begin{figure}[H]
    \centering
    \includegraphics[width=1\linewidth]{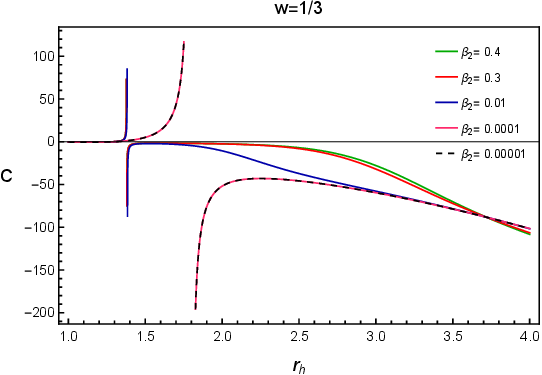}
    \caption{We plot $C$ for various $\beta_2$ values for radiation  field}
    \label{f11}
    \end{figure}

    The solution is thermodynamically stable if its specific heat (\ref{31}) is positive, indicating that the system can absorb heat without undergoing a thermal runaway.
    From Figures \ref{f10} and \ref{f11}, we observe that the specific heat is positive for small horizon radius. This indicates that a black hole becomes thermodynamically stable provided the horizon is sufficiently small. 
    \section{Conclusions}\label{sec4}
    In this study, we present a general method for solving complex differential equations arising in modified gravity theories. Additionally, we construct an approximate black hole metric within the framework of \( f(R, G, T) \) gravity and examine the
thermodynamic nature for fixed values of the equation of state parameter $w=0 ,\frac{1}{3}$. Moreover, the equations and results presented in this work are, to the best of our knowledge, established for the first time in the existing literature.
\par
\vspace{0.25cm}
(i) In this work, we have developed and demonstrated a systematic method for solving complex, nonlinear fourth-order differential (\ref{22}) equations that arise in the context of \( f(R, G, T) \) gravity. This method is particularly effective in dealing with the mathematical challenges introduced by higher-order curvature terms and non-minimal coupling to matter. By applying this approach, we have successfully derived an approximate solution (\ref{28}) for the metric function \( \psi(r) \) under the assumption of a static, spherically symmetric spacetime with an anisotropic fluid (\ref{13}) as the matter source. This approximate metric provides insight into the structure and behavior of black holes in modified gravity scenarios beyond General Relativity.
\par
\vspace{0.25cm}
(ii) By utilizing the approximate solution given in equation(\ref{28}), we have investigated the thermodynamic properties of black holes within the framework of \( f(R, G, T) \) gravity. Specifically, we analyzed the behavior of Wald’s entropy and Hawking temperature as functions of the horizon radius, as illustrated in Figures\ref{f4}-\ref{f8}. Our results show that entropy increases monotonically with the horizon radius, consistent with the expected thermodynamic behavior. In contrast, the Hawking temperature generally decreases as the horizon radius increases. 
\par
\vspace{0.25cm}

(iii) Next, we analyze the specific heat to study the thermodynamic stability of the black hole. Our analysis shows that when equation of state parameter $w=0 ,\frac{1}{3}$, the black hole exhibits thermodynamic stability only for small horizon radii. 
\par
\vspace{0.25cm}
(v) While the black holes studied in this work share several features and thermodynamic behaviors with their counterparts in Einstein and Lovelock gravity, they also exhibit unique characteristics specific to the \( f(R, G, T) \) framework. One notable distinction is the possibility of stable black holes with small horizon radius, which strongly depends on the value of the parameter \( \beta_2 \). This reveals richer solution structures that go beyond standard theories of gravity. There remains significant potential for future research in this modified gravity framework. In particular, it would be interesting to explore the effects of introducing additional matter sources, such as Maxwell fields, which could further enrich the thermodynamic behavior and lead to more complex phase structures.

\section*{Acknowledgements}
AG is thankful to IIEST, Shibpur, India, for providing an Institute
Fellowship (SRF).
\section{Data Availability Statement}
 This manuscript has no associated data.
[Author’s comment: In the present study, no datasets are generated or
analyzed]
    \bibliographystyle{naturemag}
\bibliography{bibliography}

@article{I1,
  title={Measurements of $\Omega$ and $\Lambda$ from 42 high-redshift supernovae},
  author={Perlmutter, Saul and Aldering, Goldhaber and Goldhaber, Gerson and Knop, Richard A and Nugent, Peter and Castro, Patricia G and Deustua, Susana and Fabbro, Sebastien and Goobar, Ariel and Groom, Donald E and others},
  journal={The Astrophysical Journal},
  volume={517},
  number={2},
  pages={565},
  year={1999},
  publisher={IOP Publishing}
}

@article{I2,
author = {SAHNI, VARUN and STAROBINSKY, ALEXEI},
title = {RECONSTRUCTING DARK ENERGY},
journal = {International Journal of Modern Physics D},
volume = {15},
number = {12},
pages = {2105-2132},
year = {2006},
doi = {10.1142/S0218271806009704},

URL = { 
    
        https://doi.org/10.1142/S0218271806009704
},
eprint = { 
    
        https://doi.org/10.1142/S0218271806009704
}
}

@article{I3,
  title={Reconstruction of dark energy and expansion dynamics using Gaussian processes},
  author={Seikel, Marina and Clarkson, Chris and Smith, Mathew},
  journal={Journal of Cosmology and Astroparticle Physics},
  volume={2012},
  number={06},
  pages={036},
  year={2012},
  publisher={IOP Publishing}
}

@article{I4,
  title = {Direct Reconstruction of Dark Energy},
  author = {Clarkson, Chris and Zunckel, Caroline},
  journal = {Phys. Rev. Lett.},
  volume = {104},
  issue = {21},
  pages = {211301},
  numpages = {4},
  year = {2010},
  month = {May},
  publisher = {American Physical Society},
  doi = {10.1103/PhysRevLett.104.211301},
  url = {https://link.aps.org/doi/10.1103/PhysRevLett.104.211301}
}

@article{I5,
  title={Reconstruction of dark energy and equilibrium thermodynamics in Brans-Dicke theory},
  author={Liu, Xian-Ming and Liu, Wen-Biao},
  journal={Astrophysics and Space Science},
  volume={334},
  pages={203--207},
  year={2011},
  publisher={Springer}
}

@article{I6,
title = {Unified cosmic history in modified gravity: From F(R) theory to Lorentz non-invariant models},
journal = {Physics Reports},
volume = {505},
number = {2},
pages = {59-144},
year = {2011},
issn = {0370-1573},
doi = {https://doi.org/10.1016/j.physrep.2011.04.001},
url = {https://www.sciencedirect.com/science/article/pii/S0370157311001335},
author = {Shin’ichi Nojiri and Sergei D. Odintsov},
}

@article{I7,
title = {A new type of isotropic cosmological models without singularity},
journal = {Physics Letters B},
volume = {91},
number = {1},
pages = {99-102},
year = {1980},
issn = {0370-2693},
doi = {https://doi.org/10.1016/0370-2693(80)90670-X},
url = {https://www.sciencedirect.com/science/article/pii/037026938090670X},
author = {A.A. Starobinsky},
}

@article{I8,
author = {CAPOZZIELLO, SALVATORE},
title = {CURVATURE QUINTESSENCE},
journal = {International Journal of Modern Physics D},
volume = {11},
number = {04},
pages = {483-491},
year = {2002},
doi = {10.1142/S0218271802002025},

URL = { 
    
        https://doi.org/10.1142/S0218271802002025
},
eprint = { 
       https://doi.org/10.1142/S0218271802002025    }
}

@article{I9,
  title = {Lovelock gravitational field equations in cosmology},
  author = {Deruelle, Nathalie and Fari\~na-Busto, Luis},
  journal = {Phys. Rev. D},
  volume = {41},
  issue = {12},
  pages = {3696--3708},
  numpages = {0},
  year = {1990},
  month = {Jun},
  publisher = {American Physical Society},
  doi = {10.1103/PhysRevD.41.3696},
  url = {https://link.aps.org/doi/10.1103/PhysRevD.41.3696}
}

@article{I10,
  title={The Einstein tensor and its generalizations},
  author={Lovelock, David},
  journal={Journal of Mathematical Physics},
  volume={12},
  number={3},
  pages={498--501},
  year={1971},
  publisher={American Institute of Physics}
}

@article{I11,
  title={Einsteinian cubic gravity},
  author={Bueno, Pablo and Cano, Pablo A},
  journal={Physical Review D},
  volume={94},
  number={10},
  pages={104005},
  year={2016},
  publisher={APS}
}

@article{I12,
  title = {Einstein's other gravity and the acceleration of the Universe},
  author = {Linder, Eric V.},
  journal = {Phys. Rev. D},
  volume = {81},
  issue = {12},
  pages = {127301},
  numpages = {3},
  year = {2010},
  month = {Jun},
  publisher = {American Physical Society},
  doi = {10.1103/PhysRevD.81.127301},
  url = {https://link.aps.org/doi/10.1103/PhysRevD.81.127301}
}

@article{I13,
  title = {Cosmological perturbations in $f(T)$ gravity},
  author = {Chen, Shih-Hung and Dent, James B. and Dutta, Sourish and Saridakis, Emmanuel N.},
  journal = {Phys. Rev. D},
  volume = {83},
  issue = {2},
  pages = {023508},
  numpages = {11},
  year = {2011},
  month = {Jan},
  publisher = {American Physical Society},
  doi = {10.1103/PhysRevD.83.023508},
  url = {https://link.aps.org/doi/10.1103/PhysRevD.83.023508}
}

@article{I14,
  title = {Self-gravitating spherically symmetric solutions in scalar-torsion theories},
  author = {Kofinas, Georgios and Papantonopoulos, Eleftherios and Saridakis, Emmanuel N.},
  journal = {Phys. Rev. D},
  volume = {91},
  issue = {10},
  pages = {104034},
  numpages = {14},
  year = {2015},
  month = {May},
  publisher = {American Physical Society},
  doi = {10.1103/PhysRevD.91.104034},
  url = {https://link.aps.org/doi/10.1103/PhysRevD.91.104034}
}

@article{I15,
  title = {Teleparallel equivalent of Gauss-Bonnet gravity and its modifications},
  author = {Kofinas, Georgios and Saridakis, Emmanuel N.},
  journal = {Phys. Rev. D},
  volume = {90},
  issue = {8},
  pages = {084044},
  numpages = {11},
  year = {2014},
  month = {Oct},
  publisher = {American Physical Society},
  doi = {10.1103/PhysRevD.90.084044},
  url = {https://link.aps.org/doi/10.1103/PhysRevD.90.084044}
}

@article{I16,
  title = {Cosmological applications of $F(T,{T}_{G})$ gravity},
  author = {Kofinas, Georgios and Saridakis, Emmanuel N.},
  journal = {Phys. Rev. D},
  volume = {90},
  issue = {8},
  pages = {084045},
  numpages = {10},
  year = {2014},
  month = {Oct},
  publisher = {American Physical Society},
  doi = {10.1103/PhysRevD.90.084045},
  url = {https://link.aps.org/doi/10.1103/PhysRevD.90.084045}
}

@article{I17,
  title={New Black Hole Solutions in $f(P)$ Gravity and their Thermodynamic Nature},
  author={Ghosh, Aniruddha and Debnath, Ujjal},
  journal={Physics Letters B},
  pages={139305},
  year={2025},
  publisher={Elsevier}
}

@article{I18,
  title = {Fourth order Weyl gravity},
  author = {Flanagan, \'Eanna \'E.},
  journal = {Phys. Rev. D},
  volume = {74},
  issue = {2},
  pages = {023002},
  numpages = {4},
  year = {2006},
  month = {Jul},
  publisher = {American Physical Society},
  doi = {10.1103/PhysRevD.74.023002},
  url = {https://link.aps.org/doi/10.1103/PhysRevD.74.023002}
}

@ARTICLE{I19,
       author = {{Mannheim}, Philip D. and {Kazanas}, Demosthenes},
        title = "{Exact Vacuum Solution to Conformal Weyl Gravity and Galactic Rotation Curves}",
      journal = {\apj},
         year = 1989,
        month = jul,
       volume = {342},
        pages = {635},
          doi = {10.1086/167623},
       adsurl = {https://ui.adsabs.harvard.edu/abs/1989ApJ...342..635M}
}

@article{I20,
  title = {Lovelock gravitational field equations in cosmology},
  author = {Deruelle, Nathalie and Fari\~na-Busto, Luis},
  journal = {Phys. Rev. D},
  volume = {41},
  issue = {12},
  pages = {3696--3708},
  numpages = {0},
  year = {1990},
  month = {Jun},
  publisher = {American Physical Society},
  doi = {10.1103/PhysRevD.41.3696},
  url = {https://link.aps.org/doi/10.1103/PhysRevD.41.3696}
}

@article{I21,
  title={The Einstein tensor and its generalizations},
  author={Lovelock, David},
  journal={Journal of Mathematical Physics},
  volume={12},
  number={3},
  pages={498--501},
  year={1971},
  publisher={American Institute of Physics}
}

@article{I22,
  title={Kiselev black holes in f (R, T) gravity},
  author={Santos, LCN and da Silva, FM and Mota, CE and Lobo, IP and Bezerra, VB},
  journal={General Relativity and Gravitation},
  volume={55},
  number={8},
  pages={94},
  year={2023},
  publisher={Springer}
}

@article{I23,
  title={f (R, T) gravity},
  author={Harko, Tiberiu and Lobo, Francisco SN and Nojiri, Shin’ichi and Odintsov, Sergei D},
  journal={Physical Review D—Particles, Fields, Gravitation, and Cosmology},
  volume={84},
  number={2},
  pages={024020},
  year={2011},
  publisher={APS}
}

@article{I24,
  title={Constructions of f (R, G, T) gravity from some expansions of the universe},
  author={Debnath, Ujjal},
  journal={International Journal of Modern Physics A},
  volume={35},
  number={31},
  pages={2050203},
  year={2020},
  publisher={World Scientific}
}

@article{I25,
  title={Black-hole solutions in F (R) gravity with conformal anomaly},
  author={Hendi, SH and Momeni, D},
  journal={The European Physical Journal C},
  volume={71},
  number={12},
  pages={1823},
  year={2011},
  publisher={Springer}
}

@article{I26,
  title={Charged accelerating black hole in f (R) gravity},
  author={Zhang, Ming and Mann, Robert B},
  journal={Physical Review D},
  volume={100},
  number={8},
  pages={084061},
  year={2019},
  publisher={APS}
}

@article{I27,
  title={Gravitational perturbations of a Kerr black hole in f (R) gravity},
  author={Suvorov, Arthur George},
  journal={Physical Review D},
  volume={99},
  number={12},
  pages={124026},
  year={2019},
  publisher={APS}
}

@article{I28,
  title={Regular black holes in f (G) gravity},
  author={de S. Silva, Marcos V and Rodrigues, Manuel E},
  journal={The European Physical Journal C},
  volume={78},
  pages={1--18},
  year={2018},
  publisher={Springer}
}

@article{I29,
  title={Regular multihorizon black holes in f (G) gravity with nonlinear electrodynamics},
  author={Rodrigues, Manuel E and Silva, Marcos V de S},
  journal={Physical Review D},
  volume={99},
  number={12},
  pages={124010},
  year={2019},
  publisher={APS}
}

@article{I30,
  title={Violation of the first law of black hole thermodynamics in f (T) gravity},
  author={Miao, Rong-Xin and Li, Miao and Miao, Yan-Gang},
  journal={Journal of Cosmology and Astroparticle Physics},
  volume={2011},
  number={11},
  pages={033},
  year={2011},
  publisher={IOP Publishing}
}

@article{I31,
  title={Static spherically symmetric black holes in weak f (T)-gravity},
  author={Pfeifer, Christian and Schuster, Sebastian},
  journal={Universe},
  volume={7},
  number={5},
  pages={153},
  year={2021},
  publisher={MDPI}
}

@article{I32,
  title={Exact charged black-hole solutions in D-dimensional f (T) gravity: torsion vs curvature analysis},
  author={Capozziello, Salvatore and Gonzalez, PA and Saridakis, Emmanuel N and Vasquez, Yerko},
  journal={Journal of High Energy Physics},
  volume={2013},
  number={2},
  pages={1--25},
  year={2013},
  publisher={Springer}
}

@book{F1,
  title={The classical theory of fields},
  author={Landau, Lev Davidovich},
  volume={2},
  year={2013},
  publisher={Elsevier}
}

@article{F2,
  title={Reconstruction of some cosmological models in f (R, T) cosmology},
  author={Jamil, Mubasher and Momeni, D and Raza, Muhammad and Myrzakulov, Ratbay},
  journal={The European Physical Journal C},
  volume={72},
  pages={1--6},
  year={2012},
  publisher={Springer}
}

@article{F3,
  title={Finite-time singularities in f (R, T) gravity and the effect of conformal anomaly},
  author={Houndjo, MJS and Batista, CEM and Campos, JP and Piattella, OF},
  journal={Canadian Journal of Physics},
  volume={91},
  number={7},
  pages={548--553},
  year={2013},
  publisher={NRC Research Press}
}

@article{F4,
  title={Thermodynamics in f (R, T) theory of gravity},
  author={Sharif, M and Zubair, M},
  journal={Journal of Cosmology and Astroparticle Physics},
  volume={2012},
  number={03},
  pages={028},
  year={2012},
  publisher={IOP Publishing}
}

@article{F5,
  title={f (R, T) gravity from null energy condition},
  author={Alvarenga, FG and Houndjo, MJS and Monwanou, AV and Orou, JBC and others},
  journal={Int. J. Mod. Phys},
  volume={4},
  pages={130--139},
  year={2013}
}

@article{F6,
  title={FRW cosmology in f (R, T) gravity},
  author={Myrzakulov, Ratbay},
  journal={The European Physical Journal C},
  volume={72},
  number={11},
  pages={2203},
  year={2012},
  publisher={Springer}
}

@article{F7,
  title={An alternative f (R, T) gravity theory and the dark energy problem},
  author={Chakraborty, Subenoy},
  journal={General Relativity and Gravitation},
  volume={45},
  pages={2039--2052},
  year={2013},
  publisher={Springer}
}

@article{F8,
  title={Finite-time future singularities in modified Gauss--Bonnet and $\mathcal{F}$ (R, G) gravity and singularity avoidance},
  author={Bamba, Kazuharu and Odintsov, Sergei D and Sebastiani, Lorenzo and Zerbini, Sergio},
  journal={The European Physical Journal C},
  volume={67},
  pages={295--310},
  year={2010},
  publisher={Springer}
}

@article{F9,
  title={Cosmological perturbation in f (R, G) theories with a perfect fluid},
  author={De Felice, Antonio and Gerard, Jean-Marc and Suyama, Teruaki},
  journal={Physical Review D—Particles, Fields, Gravitation, and Cosmology},
  volume={82},
  number={6},
  pages={063526},
  year={2010},
  publisher={APS}
}

@article{F10,
  title={Inevitable ghost and the degrees of freedom in f (R,) gravity},
  author={De Felice, Antonio and Tanaka, Takahiro},
  journal={Progress of Theoretical Physics},
  volume={124},
  number={3},
  pages={503--515},
  year={2010},
  publisher={Oxford University Press}
}

@article{F11,
  title={Black hole solutions surrounded by perfect fluid in Rastall theory},
  author={Heydarzade, Y and Darabi, F},
  journal={Physics Letters B},
  volume={771},
  pages={365--373},
  year={2017},
  publisher={Elsevier}
}

@article{BHT1,
  title={The four laws of black hole mechanics},
  author={Bardeen, James M and Carter, Brandon and Hawking, Stephen W},
  journal={Communications in mathematical physics},
  volume={31},
  pages={161--170},
  year={1973},
  publisher={Springer}
}

@article{BHT2,
  title = {Black Holes and Entropy},
  author = {Bekenstein, Jacob D.},
  journal = {Phys. Rev. D},
  volume = {7},
  issue = {8},
  pages = {2333--2346},
  numpages = {0},
  year = {1973},
  month = {Apr},
  publisher = {American Physical Society},
  doi = {10.1103/PhysRevD.7.2333},
  url = {https://link.aps.org/doi/10.1103/PhysRevD.7.2333}
}

@article{BHT3,
  title = {Generalized second law of thermodynamics in black-hole physics},
  author = {Bekenstein, Jacob D.},
  journal = {Phys. Rev. D},
  volume = {9},
  issue = {12},
  pages = {3292--3300},
  numpages = {0},
  year = {1974},
  month = {Jun},
  publisher = {American Physical Society},
  doi = {10.1103/PhysRevD.9.3292},
  url = {https://link.aps.org/doi/10.1103/PhysRevD.9.3292}
}

@article{BHT4,
  title = {Some properties of the Noether charge and a proposal for dynamical black hole entropy},
  author = {Iyer, Vivek and Wald, Robert M.},
  journal = {Phys. Rev. D},
  volume = {50},
  issue = {2},
  pages = {846--864},
  numpages = {0},
  year = {1994},
  month = {Jul},
  publisher = {American Physical Society},
  doi = {10.1103/PhysRevD.50.846},
  url = {https://link.aps.org/doi/10.1103/PhysRevD.50.846}
}

@article{BHT5,
  title={Particle creation by black holes},
  author={Hawking, Stephen W},
  journal={Communications in mathematical physics},
  volume={43},
  number={3},
  pages={199--220},
  year={1975},
  publisher={Springer}
}

@article{BHT6,
  title = {Black hole entropy is the Noether charge},
  author = {Wald, Robert M.},
  journal = {Phys. Rev. D},
  volume = {48},
  issue = {8},
  pages = {R3427--R3431},
  numpages = {0},
  year = {1993},
  month = {Oct},
  publisher = {American Physical Society},
  doi = {10.1103/PhysRevD.48.R3427},
  url = {https://link.aps.org/doi/10.1103/PhysRevD.48.R3427}
}

@article{BHT7,
  title = {Some properties of the Noether charge and a proposal for dynamical black hole entropy},
  author = {Iyer, Vivek and Wald, Robert M.},
  journal = {Phys. Rev. D},
  volume = {50},
  issue = {2},
  pages = {846--864},
  numpages = {0},
  year = {1994},
  month = {Jul},
  publisher = {American Physical Society},
  doi = {10.1103/PhysRevD.50.846},
  url = {https://link.aps.org/doi/10.1103/PhysRevD.50.846}
}

@article{basak2025accretion,
  title={Accretion of dark energy onto black hole in Bumblebee field},
  author={Basak, Anuka and Debnath, Ujjal},
  journal={The European Physical Journal C},
  volume={85},
  number={6},
  pages={1--17},
  year={2025},
  publisher={Springer}
}

@article{mukherjee2024accretion,
  title={Accretion Phenomena of Different Kinds of Chaplygin Gas Models onto Kehagias-Sfetsos Black Hole in Horava-Lifshitz Gravity Scenario},
  author={Mukherjee, Puja and Debnath, Ujjal and Pradhan, Anirudh},
  journal={arXiv preprint arXiv:2410.20367},
  year={2024}
}
\end{document}